\documentclass[dvipdfmx]{cs20proc}

\usepackage{kantlipsum}

\newcommand{\equref}[1]{Eq. ~(\ref{#1})}
\newcommand{\figref}[1]{Fig. ~\ref{#1}}

\newcommand{\angstrom}{\textup{\AA}} 

\usepackage{natbib}
\bibpunct[:]{(}{)}{,}{a}{}{,}

\editors{S. J. Wolk, A. K. Dupree, H. M. G\"unther}
\publisher{Zenodo}
\conference{The 20.5th Cambridge Workshop on Cool Stars, Stellar Systems, and the Sun}
\conferencedate{2021}

\title{Rotation-Activity Relation of Ca I\hspace{-.1em}I and Mg I Infrared Emission Lines of Young Stars}
\author{Mai Yamashita$^1$, Yoichi Itoh$^1$, Yuhei Takagi$^2$}

\affiliation{$^1$University of Hyogo (Japan), $^2$National Astronomical Observatory of Japan}

\shorttitle{Rotation-Activity Relation of Young Stars}
\shortauthors{Mai Yamashita, Yoichi Itoh \& Yuhei Takagi}

\abs{To reveal the detail of the internal structure, the relationship between chromospheric activity and the Rossby number, $N_{\rm R}$(= rotational period $P$ / convective turnover time $\tau_{\rm c}$), has been extensively examined for main-sequence stars. The goal of our work is to apply the same methods to pre-main-sequence (PMS) stars and identify the appropriate model of $\tau_{\rm c}$ for them. \citet{y20} investigated the relationship between $N_{\rm R}$ and strengths of the Ca I\hspace{-.1em}I infrared triplet (IRT; $\lambda 8498, 8542, 8662 $ $\, \mathrm{\angstrom}$) emission lines of 60 PMS stars. Their equivalent widths are converted into the emission line to stellar bolometric luminosity ratio ($R^{\prime}$). The 54 PMS stars have $N_{\rm R} < 10^{-1.0}$ and show $R^{\prime} \sim 10^{-4.2}$ as large as the maximum $R^{\prime}$ of the zero-age main-sequence (ZAMS) stars. However, because all $R^{\prime}$ was saturated against $N_{\rm R}$, it was not possible to estimate the appropriate $\tau_{\rm c}$ model for the PMS stars. We noticed that Mg I emission lines at $8808 \, \mathrm{\angstrom}$ is an optically thin chromospheric line, appropriate for determination of the adequate $\tau_{\rm c}$ for PMS stars. Using the archive data of the Anglo-Australian Telescope (AAT)/the University College London Echelle Spectrograph (UCLES), we investigated the Mg I line of 52 ZAMS stars. After subtracting photospheric absorption component, the Mg I line is detected as an emission line in 45 ZAMS stars, whose $R^{\prime}_{\rm Mg I}$ is between $10^{-5.9}$ and $10^{-4.1}$. The Mg I line is not saturated yet in "the saturated regime for the Ca I\hspace{-.1em}I emission lines", i.e. $10^{-1.6} < N_{\rm R} < 10^{-0.8}$. Therefore, the adequate $\tau_{\rm c}$ for PMS stars can be determined by measuring of their $R^{\prime}_{\rm Mg I}$ values. }

\begin{document}

\maketitle

\section{Introduction}
Chromosphere is the region between photosphere and corona. The temperature of the chromosphere gradually increases with radial distance to the photosphere. In an active chromospheric region, atoms emit permitted lines such as H$\rm \alpha$ and Ca I\hspace{-.1em}I. It is claimed that chromospheric activity is driven by the magnetic field, which is generated by the dynamo process. 

In the dynamo process, the Coriolis force (= rotational moment $\times$ convection velocity) balances with the Lorentz force (= current $\times$ magnetic strength / density of plasma) (Baliunas et al. 1996). It is widely considered that a fast rotating star has strong dynamo, thus showing strong chromospheric emission lines. Because pre-main sequence (PMS) stars are fast rotators (Gallet \& Bouvier 2015), they are expected to have high chromospheric activities. Actually, \citet{y20} confirmed strong chromospheric emission lines for PMS stars. 

To reveal the structure of the convection zone, the relationship between chromospheric line strengths and the Rossby number, 
\begin{equation}
N_{\rm R} \equiv \frac{\mbox{rotational period } P}{\mbox{convective turnover time } \tau_{\rm c}}, 
\end{equation}
has been extensively examined for main-sequence stars. Noyes et al. (1984) constructed the adequate $\tau_{\rm c}$ model with $\alpha (\equiv \mbox{mixing length} \, l / \mbox{pressure scale height} \, H_p) = 1.9$. 


$\alpha$ is one of the most critical parameters on the calculation of evolutionary tracks. \citet{jk07} assumed $\alpha = 0.5$ and constructed the evolutionary tracks, whereas \citet{dm94} constructed four evolutionary tracks with $\alpha = 0.5, 1.2, 1.5, 2.0$, respectively. Different $\alpha$ provides different evolutionary tracks. For example, a PMS star is identified as ($1.0 \, \mathrm{M_{\odot}}, 3 \, \mathrm{Myr}$) by an evolutionary track with $\alpha=0.5$, despite ($1.2 \, \mathrm{M_{\odot}}$, $1 \, \mathrm{Myr}$) by that with $\alpha=2.0$. $\alpha$ is one of the most fundamental but unexplored parameters on stellar structure.

The goal of our work is to apply the Noyes' method to PMS stars and identify the appropriate model of $\tau_{\rm c}$. PMS stars (CTTS and WTTS) have both shorter rotation period, $P$ \citep{ga13} and longer convective turnover time, $\tau_{\rm c}$ \citep{jk07}. Thus it is expected that PMS stars have small $N_{\rm R}$ and active chromosphere.

\section{Observations and Data Reduction}
In this study, the Ca I\hspace{-.1em}I infrared triplet emission lines (IRT; $\lambda 8498, 8542, 8662 \, \mathrm{\angstrom}$) are investigated for 60 PMS stars
($0.035-2.5 \, \mathrm{M_{\odot}}, 10^5 - 10^8 \, \mathrm{yr}$), being associated with four molecular clouds or five moving groups; the Taurus-Auriga molecular cloud, the Orionis OB 1c association, the Upper Scorpius association, the Perseus molecular cloud, the TW Hydrae association, the $\eta$ Chamaeleontis cluster, the ''Cha-Near'' region, the $\beta$ Pictoris moving group, and the AB Doradus moving group. The medium- and high-resolution spectroscopy were conducted with $2 \, \mathrm{m}$ Nayuta Telescope/MALLS at Nishi-Harima Astronomical Observatory and Subaru/HDS. Archival data obtained with the Keck/HIRES, VLT/UVES, and VLT/X-Shooter were also used. A detailed description of the data reduction methods used here is presented in \citet{y20}.

We also investigated the Mg I line of F, K, G type 52 ZAMS stars by using $3.9 \, \mathrm{m}$ Anglo-Australian Telescope (AAT)/the University College London Echelle Spectrograph (UCLES) archive spectra. The PI is S. C. Marsden, date of the observations are 2000-03-17, 18, 19, 2001-01-06, 07, 08, and 2001-02-11, 12. The wavelength coverage was between $3522 \, \mathrm{\angstrom}$ and $9386 \, \mathrm{\angstrom}$. The integration time for each object was between 300 s and 1200 s. A detailed description of the observations is presented in \citet{m09}. We selected the objects from single stars in IC 2391 \cite[$30 \pm 5 \, \mathrm{Myr}$; ][]{ba04} and IC 2602 \cite[$50 \pm 5 \, \mathrm{Myr}$; ][]{st97} clusters, whose membership was confirmed by \citet{m09}. 

\begin{figure}[htb]
	\centering
	\includegraphics[width=0.9\linewidth]{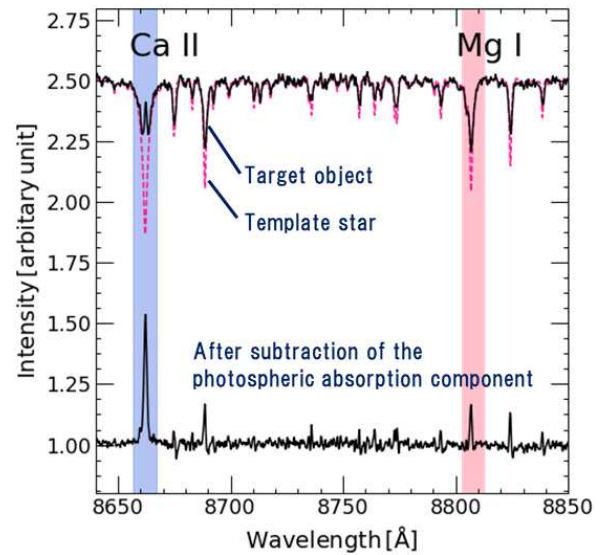} \caption{The procedures of the spectral subtaction of the photospheric component for the PMS star, RECX 11. The observed spectrum of RECX 11 is shown in the top of the panel with a solid line. The dotted line is the spectrum of fitted inactive template star. The difference between the observed and template spectra is shown in the bottom of the panel, where the Ca I\hspace{-.1em}I IRT narrow line and the Mg I narrow line appear in emission. The observed spectrum of RECX 11 and that of the template star are shown shifted by $+1.5$ for display purposes.} \label{fig:recx11}
\end{figure}

We used the Image Reduction and Analysis Facility (IRAF) software package\footnote{IRAF is distributed by the National Optical Astronomy Observatories, which are operated by the Association of Universities for Research in Astronomy, Inc., under cooperative agreement with the National Science Foundation.} for data reduction. Overscan subtraction, bias subtraction, flat fielding, removal of scattered light, removal of cosmic rays, extraction of a spectrum, wavelength calibration using a Tr-Ar lamp, continuum normalization, combining spectra, and subtracting the photospheric absorption component were conducted for all the spectra obtained by UCLES. The Mg I emission component is always buried by the photospheric absorption as shown in \figref{fig:recx11}. 
In this study, inactive stars with a spectral type similar to that of the target were used as template stars. We reduced UCLES spectra of HD 16673 (F6V), $\rm \alpha$ Cen A (G2V), $\rm \alpha$ Cen B (K1V). We also obtained the VLT archive spectra of ten inactive stars choosen from the inactive stars library \citep{yee} and UVES POP, namely $\zeta$ Ser (F2IV), HD 3861 (F8V), HD 1388 (G0V),  HD 156846 (G1V), HD 109749 (G3V), HIP 94256 (G5V), HD 190360 (G7V), HD 217107 (G8V), HD 16160 (K3V), and HD 165341B (K4V). As a result of selecting inactive stars having similar metalicity with IC 2391 \cite[$0.00\pm 0.01$; ][]{ra01} and IC 2602 \cite[$-0.01\pm 0.02$; ][]{d09}, their metalicity is $-0.11 \leq {\rm [Fe/H]} \leq 0.31$. For the correction of the rotational broadening, the spectra of the template stars were convolved with a Gaussian kernel to match the width of the absorption lines of each object. By using high $S/N$ spectra of inactive field stars, we carefully subtracted the photospheric absorption component. We considered that the subtraction was reliable if nearby faint photospheric absorption lines, such as Ni I ($6644, 6728, 7525\, \mathrm{\angstrom}$), Mn I ($6014, 6017, 6022\, \mathrm{\angstrom}$), V I  ($6039\, \mathrm{\angstrom}$), and Ti I ($7523\, \mathrm{\angstrom}$), disappeared in the spectrum after the subtraction. Finally, we measured the equivalent widths (${\rm EQW}$) of the Mg I emission line. Before measuring the EQWs, the continuum component of the spectra was added to unity. To obtain the EQWs of the Mg I emission lines, the area of the emission profile was directly integrated. The typical EQW errors were estimated by multiplying the standard deviation of the continuum by the wavelength range of the Mg I emission line of the five ZAMS stars. Avoiding any emission and absorption lines, the standard deviation of the continuum was determined by averaging over the value in $\lambda 8798-8802 \, \mathrm{\angstrom}$ and $\lambda 8813-8819 \, \mathrm{\angstrom}$. 

\section{Results}
Seven PMS stars have broad emission lines of Ca I\hspace{-.1em}I $\lambda 8498 \, \mathrm{\angstrom}$ (FWHM $> 100 \, \mathrm{km \cdot s^{-1}}$), while most PMS stars exhibit narrow emission lines (FWHM $\leq 100 \, \mathrm{km \cdot s^{-1}}$). The emission lines of DG Tau, DL Tau, and DR Tau are broad and strong (${\rm EQW} \sim 50 \, \mathrm{\angstrom}$), while those of RY Tau, SU Aur, RECX 15, and RX J1147.7-7842 are broad but not strong (${\rm EQW} < 10 \, \mathrm{\angstrom}$). All EQWs of the narrow emission lines are weaker than $5 \, \mathrm{\angstrom}$. 

\begin{figure}[htb]
	\centering
	\includegraphics[width=0.75\linewidth]{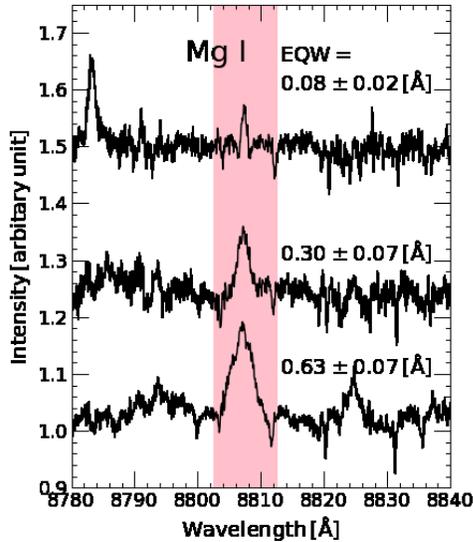}
	\caption{Examples of the emission profiles of the Mg I line at $8807.8 \, \mathrm{\angstrom}$. Photospheric absorption lines have already been subtracted.} \label{resultEQW}
\end{figure}

\figref{resultEQW} shows the Mg I emission line at $\lambda 8808 \, \mathrm{\angstrom}$ of ZAMS stars after subtracting the photospheric absorption. Before subtracting the photospheric absorption component, the Mg I emission component is always buried by the photospheric absorption. As a result of the data reduction, the Mg I line is detected as an emission line in 45 ZAMS stars. The Mg I line of these ZAMS stars shows narrow emission, indicative of chromospheric origin. Their EQWs range from $0.02 \, \mathrm{\angstrom}$ to $0.63 \, \mathrm{\angstrom}$. The Mg I emission lines of 33 ZAMS stars have EQWs less than $0.1 \, \mathrm{\angstrom}$. 5 ZAMS stars (Cl* IC2391 L32, VXR PSPC 03A, VXR PSPC 16A, VXR PSPC 45A, {[}RSP95{]} 85) do not show the Mg I line in emission or absorption, but like simply continuum component. 

\section{Discussion}
\subsection{Chromospheric activity and mass accretion rate}

We compared the strengths of the Ca I\hspace{-.1em}I IRT emission lines and mass accretion rates to discuss whether the chromosphere is activated by mass accretion from the protoplanetary disk.
\citet{mo05} investigated the chromospheric activity of CTTSs, very low-mass young stars ($0.075 \leq M_* <  0.15 \, \mathrm{M_{\odot}}$) and young brown dwarfs ($M_* \leq 0.075 \, \mathrm{M_{\odot}}$). They selected ''accretors'' by applying a number of accretion diagnostics, such as an H${\rm \alpha} 10 \%$ width $\geq 200 \, \mathrm{km \cdot s^{-1}}$. For the accretors, the surface flux of the Ca I\hspace{-.1em}I emission line, $F^{\prime}_{\lambda 8662}$, showed a positive correlation with their mass accretion rate, $\dot{M}$, for approximately 4 orders of magnitude. Hence, they claimed that the mass accretion rate can be estimated using the strength of the broad components of the Ca I\hspace{-.1em}I IRT emission lines.

In our work, $\dot{M}$ were taken from several studies, which were estimated from the amount of veiling. For objects whose veiling had not been determined, we referred $\dot{M}$ estimated with an EQW of the H$\mathrm{\alpha}$ emission line. The EQWs of the Ca I\hspace{-.1em}I emission lines are converted into $F^{\prime}$. The method of calculating $F^{\prime}$ is described in \citet{y20}.

\begin{figure}[htb]
	\centering
	\includegraphics[width=0.99\linewidth]{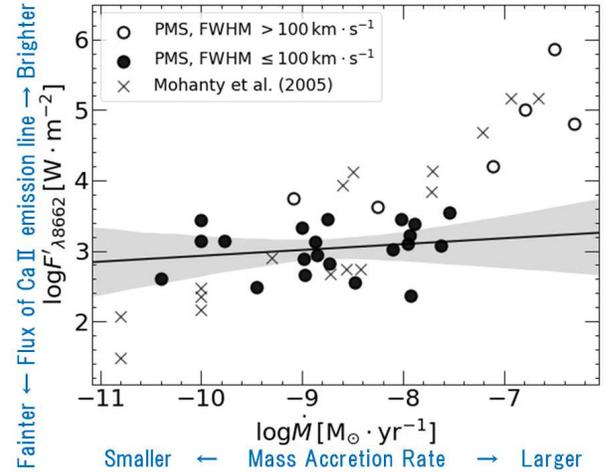}
	\caption{Surface flux of the Ca I\hspace{-.1em}I emission line, $F^{\prime}_{\lambda 8662}$, as a function of mass accretion rate, $\dot{M}$. The open circles show the objects with a broad Ca I\hspace{-.1em}I IRT emission line (FWHM $> 100 \, \mathrm{km \cdot s^{-1}}$) and the filled circles represent the objects with a narrow Ca I\hspace{-.1em}I IRT emission line (FWHM $\leq 100 \, \mathrm{km \cdot s^{-1}}$). The cross symbols show the CTTSs, very low-mass young stars ($0.075 \leq M_* <  0.15 \, \mathrm{M_{\odot}}$) and young brown dwarfs ($M_* \leq 0.075 \, \mathrm{M_{\odot}}$) studied in \citet{mo05}. } \label{figM}
\end{figure}

We see two groups among the PMS stars (\figref{figM}). The PMS stars showing a broad Ca I\hspace{-.1em}I IRT emission line have high mass accretion rate ($\dot{M} \gtrsim 10^{-7} \, \mathrm{M_{\odot} \cdot yr^{-1}}$). Those are classified as accretors based on the diagnosis of \citet{mo05}. On the other hand, the PMS stars showing a narrow emission have $\log F^{\prime}_{\lambda 8662} \sim 3$ with a flat distribution to $\dot{M}$. However, we note that the Ca I\hspace{-.1em}I IRT emission lines of PMS stars vary with time. In this work, BP Tau shows narrow emissions (${\rm EQW} = 1.15 \, \mathrm{\angstrom}$), whereas showed broad emission features (${\rm EQW} = 7.8 \, \mathrm{\angstrom}$) in \citet{mo05}.

\subsection{Rotation-Activity Relation of Ca I\hspace{-.1em}I emission lines}
\begin{figure}[htb]
	\centering
	\includegraphics[width=0.99\linewidth]{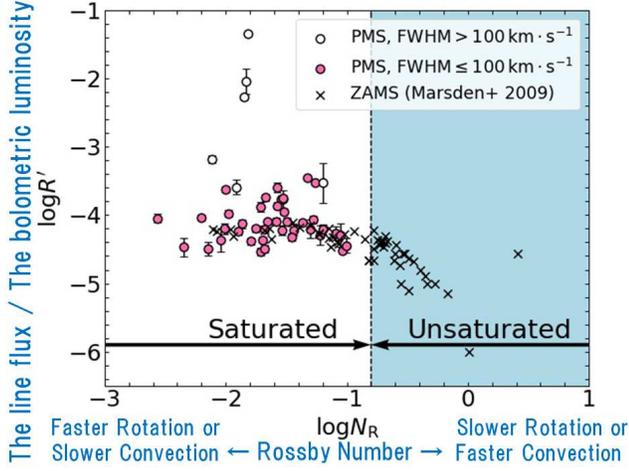}
	\caption{Relationship between the ratio of the surface flux of the Ca I\hspace{-.1em}I IRT line to the stellar bolometric luminosity, $R^{\prime}$ and the Rossby number, $N_{\rm R}$ \citep{y20}. The filled circles represent the PMS stars with narrow Ca I\hspace{-.1em}I IRT emission lines and the open circles show the PMS stars with broad Ca I\hspace{-.1em}I IRT emission lines. The cross symbols represent ZAMS stars in the young open clusters IC 2391 and IC 2602 \citep{m09}.} \label{fig:CaII}
\end{figure}

We examined the relationship between the ratio of the surface flux of the Ca I\hspace{-.1em}I IRT emission line to the stellar bolometric luminosity, $R^{\prime}$, and $N_{\rm R}$ of the 60 PMS stars (\figref{fig:CaII}). For calculating $N_{\rm R}$, we estimated $\tau_{\rm c}$ of the PMS stars using pre-main sequence evolutionary tracks presented in \citet{jk07}. 
In this track, $\tau_{\rm c}$ of low-mass stars near the main sequence was not calculated. For such objects, we applied the approximation of \citet{noyes}. A detailed description of the calculation of $N_{\rm R}$ and $R^{\prime}$ is presented in \citet{y20}. 

Only three PMS stars show broad and strong emissions, indicative of large mass accretion. Most of the PMS stars have narrow and weak emissions. This indicates that their chromospheric activity is induced by the dynamo process. All of their Ca I\hspace{-.1em}I IRT emission lines have $R^{\prime} \sim 10^{-4.2}$, which is as large as the maximum $R^{\prime}$ of ZAMS stars \citep{m09}. The PMS stars have $N_{\rm R} < 10^{-0.8}$ and constant $R^{\prime}$ against $N_{\rm R}$; i.e., the Ca I\hspace{-.1em}I IRT emission lines of the PMS stars are saturated. The chromosphere of these stars is considered to be completely filled by the active region. 

Because all $R^{\prime}$ of Ca I\hspace{-.1em}I IRT emission lines were saturated against $N_{\rm R}$, it was not possible to estimate the appropriate $\tau_{\rm c}$ model for the PMS stars. PMS stars with unsaturated Ca I\hspace{-.1em}I IRT emission lines should be investigated. The Ca I\hspace{-.1em}I IRT emission lines are not saturated for the ZAMS stars with $N_{\rm R} > 10^{-0.8}$. However no PMS star is expected to have $N_{\rm R} > 10^{-0.8}$ due to their fast rotation and long $\tau_{\rm c}$. 

As a consequence, other optically thin chromospheric emission line should be observed. As claimed in \citet{l17}, the column densities of Ca I\hspace{-.1em}I are very large in late-type stars. Moreover, Ca I\hspace{-.1em}I H, K lines heavily obscured by interstellar medium due to their short wavelength. In addition to this, their large optical depth are pointed out in \citet{hp92}. In case of He I emission line, it is difficult to discuss where is the emitting region. Mg II h, k lines are expected to blend each other in PMSs because they often rotate fast \citep{ne95}.

\subsection{Rotation-Activity Relation of Mg I emission lines}
We noticed that Mg I line at $8808 \, \mathrm{\angstrom}$ is a relatively optically thin and isolated chromospheric line. This Mg I emission line has ever been observed only for one object, the Sun. The solar observations suggest that it forms in the chromosphere about $500 \, \mathrm{km}$ above the photosphere \citep{f94}. 

\begin{figure}[htb]
	\centering
	\includegraphics[width=0.99\linewidth]{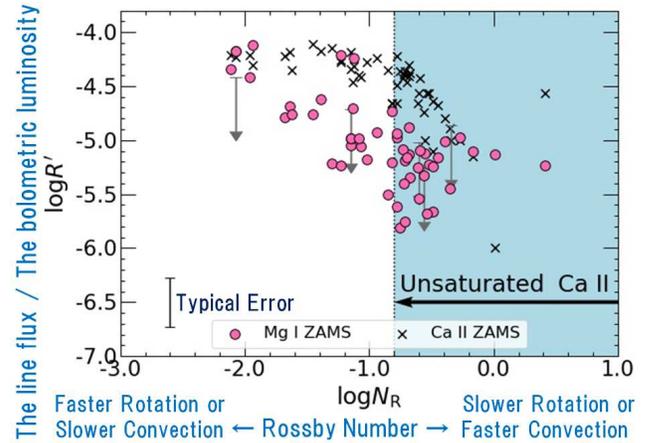}
	\caption{The relationship between the ratio of the surface flux of the Mg I line ($\lambda 8807.8 \, \mathrm{\angstrom}$) to the stellar bolometric luminosity, $R^{\prime}_{\rm Mg I}$ and the Rossby number, $N_{\rm R}$ , of ZAMS stars (filled circles). The cross symbols represent $R^{\prime}$ of Ca I\hspace{-.1em}I IRT emission lines of the ZAMS stars \citep{m09}.} \label{mgR2}
\end{figure}

We examined the relationship between the ratio of the surface flux of the Mg I emission line to the stellar bolometric luminosity, $R^{\prime}_{\rm Mg I}$, and $N_{\rm R}$ of the ZAMS stars (\figref{mgR2}). In this work, we refereed $N_{\rm R}$ calculated by \citet{m09}. They used $P$ for 21 ZAMS stars or $v \sin i$ for 31 ZAMS stars into \equref{rossbym09}, and calculated $N_{\rm R}$.
\begin{equation}
\label{rossbym09}
N_{\rm R} \equiv \frac{P}{\tau_{\rm c}} = \frac{2 \pi R_*}{\tau_{\rm c} v \sin i}<\sin i>
\end{equation}
We calculated the ratio of the surface flux of the emission line to the stellar bolometric luminosity, $R^{\prime}$, for the Mg I emission line. For calculating the surface flux $F^{\prime}$, a bolometric continuum flux per unit area at a stellar surface, $F$, was calculated at first. We used the $i$-band mag (the AB system) of the UCAC4 Catalogue \citep{za13}. For nine objects, namely VXR PSPC 45A, VXR PSPC 80A, [RSP95] 10, [RSP95] 15A, [RSP95] 35, [RSP95] 43, [RSP95] 45A, [RSP95] 79, [RSP95] 8A, [RSP95] 88A, their $i$-band mag are not listed in the UCAC4 Catalogue \citep{za13}, so that we used the $i$-band mag listed in \citet{m09}. $F$ is given as
\begin{eqnarray}
	\log \frac{f}{f_0} & = & - \frac{2}{5} \times m_{i*},\\
	F & = & f \times \left( \frac{d}{R_*} \right)^2,
\end{eqnarray}
where $f$ is the bolometric continuum flux of the object per unit area as observed on the Earth. $m_{i*}$ is the apparent magnitude of the object in the $i$-band. The bolometric continuum flux per unit area under $m_i = 0 \, \mathrm{mag}$ (the AB system) condition, $f_0$, is $1.852 \times 10^{-12} \, \mathrm{W \cdot m^{-2} \cdot \angstrom^{-1}}$ \citep{f96}. $d$ denotes the distance of an object from the Earth \citep[{\it Gaia} DR2: ][]{ba18}. Unfortunately, the distance of VXR PSPC 50A, VXR PSPC 80A, [RSP95] 10, [RSP95] 15A, [RSP95] 45A, [RSP95] 88A, [RSP95] 8A, and [RSP95] 42C are not listed in {\it Gaia} DR2, so we substituted that of IC 2391 and IC 2602 \citep[$145 \, \mathrm{pc}$; ][]{v07}. $R_*$ is the stellar radius estimated using Stefan-Boltzmann's law with the photospheric luminosity, and $T_{\rm eff}$ of the objects in {\it Gaia} DR2. For object whose $T_{\rm eff}$ is not listed in {\it Gaia} DR2, we used the luminosity and $T_{\rm eff}$ listed in \citet{m09}; VXR PSPC 50A, VXR PSPC 80A, [RSP95] 10, [RSP95] 15A, [RSP95] 45A, [RSP95] 88A, [RSP95] 8A, and Cl* IC2602 W79. $F$ was multiplied by the EQW of the Mg I emission lines.
\begin{equation}
	F^{\prime} = F \times {\rm EQW}, 
\end{equation}
For calculating $R^{\prime}$, $F^{\prime}$ are divided by $\sigma T_{\rm eff}^4$.
\begin{equation}
	R^{\prime} = \frac{F^{\prime} }{\sigma T_{\rm eff}^4}, 
\end{equation}
where $\sigma$ is Stefan-Boltzmann's constant. The dependence of the surface flux upon the $T_{\rm eff}$ of the objects is eliminated by this calculation. The typical uncertainty of $R^{\prime}$ shown in \figref{mgR2} is estimated with the error of EQW of five ZAMS stars, ignoring the uncertainties of other parameters such as $T_{\rm eff}$.

We found that the Mg I line is still not saturated in the region of $10^{-1.8} < N_{\rm R} < 10^{-0.8}$, which is the saturated region for Ca I\hspace{-.1em}I IRT emission lines. According to the discussion by \citet{m09}, the chromospheric saturation is suggested to be caused by the similar to coronal saturation. Suppose if the emitting region of Ca I\hspace{-.1em}I are completely filled, the unsaturation of the Mg I emission line suggests that the area of the Mg I emitting region may be smaller than that of Ca I\hspace{-.1em}I. Otherwise Mg I is optically thinner than Ca I\hspace{-.1em}I. We expected that the Mg I line is useful for estimating appropriate $\tau_{\rm c}$ of PMS stars.

\section{Conclusion}
\begin{enumerate}
  \item Seven PMS stars (DG Tau, DL Tau, DR Tau, RY Tau, SU Aur, RECX 15, and RX J1147.7-7842) have broad Ca I\hspace{-.1em}I IRT emission lines. It is suggested that their broad emissions result from heavy mass accretion from their protoplanetary disks.

  \item Most PMS stars have narrow Ca I\hspace{-.1em}I IRT emission lines similar to zero-age main-sequence stars in young open clusters. The emissions of these objects indicate no correlation with the mass accretion rate. The ratio of the surface flux of the Ca I\hspace{-.1em}I IRT emission lines to the stellar bolometric luminosity, $R^{\prime}$, of these objects are as large as the largest $R^{\prime}$ of the cluster members. Most PMS stars have chromospheric activity similar to zero-age main-sequence stars. The chromosphere of these stars are completely filled by the Ca I\hspace{-.1em}I emitting region.

  \item The Mg I line is not saturated yet in "the saturated regime for the Ca I\hspace{-.1em}I emission lines", i.e. $10^{-1.6} < N_{\rm R} < 10^{-0.8}$. The adequate $\tau_{\rm c}$ for PMS stars can be determined by measuring of their $R^{\prime}_{\rm Mg I}$ values. 
\end{enumerate}

\section*{Acknowledgments}
{This research has made use of the Keck Observatory Archive (KOA), which is operated by the W. M. Keck Observatory and NASA Exoplanet Science Institute (NExScI), and it is under contract with the National Aeronautics and Space Administration, and is based on the observations made with ESO Telescopes at the La Silla Paranal Observatory under programmes ID 075.C-0321, 082.C-0005, 084.C-1095, 085.C-0238, 086.C-0173, 094.C-0327, 094.C-0805, and 094.C-0913. The researcher was supported by a scholarship from the Japan Association of University Women, and I would like to appreciate them. Finally, as special thanks, M. Yamashita would also like to express our gratitude and eternal love to our cat, Miiko, who passed away on April 6, 2021. } 

\bibliographystyle{cs20proc}

\begin{thebibliography}{99}
	\bibitem[Bailer-Jones et al. (2018)]{ba18} Bailer-Jones, C. A. L., Rybizki, J., Fouesneau, M., Mantelet, G., \& Andrae, R. 2018, ApJ., 156, 58
	\bibitem[Baliunas et al.(1996)]{ba96} Baliunas, S., Sokoloff, D., \& Soon, W. 1996, ApJ., 457, L99
	\bibitem[Barrado y Navascues et al.(2004)]{ba04} Barrado y Navascues, D., Stauffer, J. R., \& Jayawardhana, R. 2004, Astrophys. J., 614, 386
	\bibitem[D'Antona \& Mazzitelli(1994)]{dm94} D'Antona, F., \& Mazzitelli, I. 1994, ApJS, 90, 467
	\bibitem[D'Orazi \& Randich(2009)]{d09} D'Orazi, V., \& Randich, S. 2009, Astron. Astrophys., 501, 553
	\bibitem[Fleck et al.(1994)]{f94} Fleck, B., Deubner, F.-L., Maier, D., \& Schmidt, W. 1994, IAU Symp, 154, 65
	\bibitem[Fukugita et al. (1996)]{f96} Fukugita, M., Ichikawa, T., Gunn, J. E., et al. 1996, Astron. J., 111, 1748 
	\bibitem[Gaia Collaboration (2018)]{g18} Gaia Collaboration, 2018, Astron. Astrophys., 616, A1
	\bibitem[Gallet \& Bouvier (2013)]{ga13} Gallet, F., \& Bouvier, J. 2015, A\&A, 577, 1
	\bibitem[Hamann \& Persson (1992)]{hp92} Hamann, F., \& Persson, S. E. 1992a, Astrophys. J. Suppl., 82, 247
	\bibitem[Jung \& Kim(2007)]{jk07} Jung, Y. K., \& Kim, Y.-C. 2007, J. Astron. Space Sci., 24, 1
	\bibitem[Linsky (2017)]{l17} Linsky, J. L. 2017, Annu Rev Astron Astrophys, 55, 159
	\bibitem[Marsden et al.(2009)]{m09} Marsden, S. C., Carter, B. D., \& Donati, J.-F. 2009, MNRAS, 399, 888
	\bibitem[Mohanty et al.(2005)]{mo05} Mohanty, S., Jayawardhana, R., \& Basri, G. 2005, ApJ, 626, 498
	\bibitem[Neuhauser et al. (1995)]{ne95} Neuhauser, R., Strerzik, M. F., Schmitt, J. H. M. M., Wichmann, R., \& Krautter, J. 1995, Astron. Astrophys., 297, 391
	\bibitem[Noyes et al.(1984)]{noyes} Noyes, R. W., Hamann, F. W., Baliunas, S. L., \& Vaughan, A. H. 1984, AJ, 279, 763
	\bibitem[Patten \& Simon (1996)]{ps96} Patten, B. M., \& Simon, T. 1996, Astrophys. J. Suppl., 106, 489
	\bibitem[Randich et al. (1995)]{ra95} Randich, S., Schmitt, J. H. M. M., Prosser, C. F., \& Stauffer, J. R. 1995, Astron. Astrophys., 300, 134
	\bibitem[Randich et al. (2001)]{ra01} Randich, S., Pallavincini, R., Meola, G., Stauffer, J. R., \& Balachandran, S. 2001, Astron. Astrophys., 372, 862
	\bibitem[Stauffer et al. (1997)]{st97}Stauffer, J. R., Hartmann, L. W., Prosser, C. F., et al. 1997, Astrophys. J., 479, 776
	\bibitem[van Leeuwen (2007)]{v07} van Leeuwen, F. 2007, Astron. Astrophys., 474, 653
	\bibitem[Yee, Petigura \& von Braun (2017)]{yee} Yee, S. W., Petigura, E. A., \& von Braun, K. 2017, Astrophys. J., 836, 77
	\bibitem[Yamashita et al.(2020)]{y20} Yamashita, M., Itoh, Y., \& Takagi, Y. 2020, PASJ, 72, 80
	\bibitem[Zacharias et al. (2013)]{za13} Zacharias, N., Finch, C. T., Girard, T. M., et al. 2013, Astrophys. J., 145, 1
\end{thebibliography}

\end{document}